\documentclass[aps,pr2,twocolumn,superscriptaddress,showpacs]{revtex4}

\usepackage{graphicx, latexsym, verbatim}
\usepackage{amssymb, amsmath}
\usepackage[ansinew]{inputenc}
\usepackage{color}
\usepackage[english]{babel}

\newcommand{\bEqa}{\begin{eqnarray}}
\newcommand{\eEqa}{\end{eqnarray}}

\newcommand{\Ub}{\mathbf{U}}

\begin{document}

\title{Electrochemical control of quantum interference in anthraquinone-based molecular switches}

\author{Troels Markussen}
\email[]{trma@fysik.dtu.dk}
\affiliation{Center for Atomic-scale Materials Design (CAMD), Department of Physics, Technical University of Denmark, DK-2800 Kgs. Lyngby, Denmark}
\affiliation{Danish National Research Foundations Center of Individual Nanoparticle Functionality (CINF), Department of Physics, Technical University of Denmark, DK-2800 Kgs. Lyngby, Denmark}

\author{Jakob Schi\"otz}
\affiliation{Danish National Research Foundations Center of Individual Nanoparticle Functionality (CINF), Department of Physics, Technical University of Denmark, DK-2800 Kgs. Lyngby, Denmark}

\author{Kristian S. Thygesen}
\affiliation{Center for Atomic-scale Materials Design (CAMD), Department of Physics, Technical University of Denmark, DK-2800 Kgs. Lyngby, Denmark}




\begin{abstract}
Using first-principles calculations we analyze the electronic transport properties of a recently proposed anthraquinone based electrochemical switch. Robust conductance on/off ratios of several orders of magnitude are observed due to destructive quantum interference present in the anthraquinone, but absent in the hydroquinone molecular bridge. A simple explanation of the interference effect is achieved by transforming the frontier molecular orbitals into localized molecular orbitals thereby obtaining a minimal tight-binding  model describing the transport in the relevant energy range in terms of hopping via the localized orbitals.  The topology of the tight-binding model, which is dictated by the symmetries of the molecular orbitals, determines the amount of quantum interference. 
\end{abstract}

\maketitle

\section{Introduction}
Molecular electronics is a very active area of research which holds the promise of continuing the miniaturization of active devices beyond the limits of standard silicon technologies~\cite{JoachimNature2000,Weibel2007,poulsen09}. Of particular interest are molecules which can be switched between two distinct states with different electronic conductance~\cite{SenseJanReview}. Reversible switching between the 'on' and 'off' states can be mediated with light~\cite{Yasuda2003,Kronemeijer2008, Molen2009,Kudernac2009}, bias voltage~\cite{Loertscher2006}, or by changes in the electrochemical environment~\cite{Gittins2000,Haiss2003}. It has recently been proposed by van Dijk et al.~\cite{DijkOrgLett2006} to use an anthraquinone based molecule as an electrochemical switch. Cyclic voltametry showed that the anthraquinone (AQ) could be reversibly reduced to hydroquinone (HQ), with a significantly different UV-vis absorption spectra.  In this communication we calculate the conductance properties of the proposed switch and show that very large on/off ratios should be expected due to destructive quantum interference effects in the off-state (AQ) molecule.

It has previously been shown that quantum interference effects can have a dramatic impact on the electronic transport through a single molecule~\cite{Baer2002,Stadler2004,Cardamone2006,Ke2008}. As an example, experiments have shown that benzene has very different transport properties depending on the position of the anchor groups~\cite{Patoux1997,Mayor2003}. This has been confirmed by theoretical works~\cite{Hansen2009} and explained in terms of interfering pathways. Interference effects in benzene have also been studied for the non-coherent coulomb blockade regime~\cite{Begemann2008,Darau2009}. Related theoretical studies has recently been concerned with interference effects in nitrobenzene~\cite{StadlerPRB2009} and in cross conjugated molecules~\cite{Andrews2008,Solomon2009} showing transmission anti-resonances close to the Fermi energy, thus potentially making the destructive quantum interference features useful in molecular electronics applications. Contrary to the previously discussed cross conjugated molecules, the AQ-HQ setup offers the possibility to electrochemically switch between a situation with and without transmission anti-resonances.  

We show in this work that the interference effects in anthraquinone electrochemical switches can be understood from a minimal tight-binding (TB) model constructed from the frontier molecular orbitals of the free molecule. Such simple models are clearly desirable for an intuitive understanding and for a fast screening of molecules exhibiting interesting interference features.

\section{DFT-Landauer transmissions}
\begin{figure}[htb!]   
	\includegraphics[width=\columnwidth,angle=0]{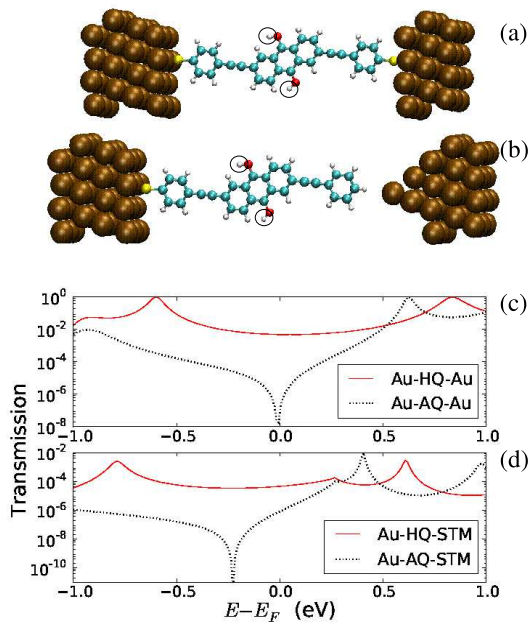}
\caption{Atomic structure of HQ molecule coupled to gold electrodes (a). The two hydrogens in the circles can reversibly be attached or removed electrochemically forming HQ (with H-atoms) and AQ (without H-atoms). Panel (b) shows an STM setup where the molecule is bound to only one Au surface. Panels (c) and (d) show the electronic transmission functions for the two molecular contacts in panel (a) and (b), respectively. Solid curves correspond to HQ while dashed curves show AQ results. Both setup give large on-off ratios at the Fermi level.}
\label{thio_trans}
\end{figure}

We calculate the electronic conductance of the molecular contacts shown in Figure \ref{thio_trans} (a)-(b), using a combination of density functional theory (DFT) and Green's function (GF) methods. In Figure \ref{thio_trans} (a) the molecule is terminated with two S-atoms that bind to both Au surface. This contact resembles the situation in e.g. mechanical break junction experiments. Figure \ref{thio_trans} (b) represents an STM experimental situation, where the molecule is only bound to one Au surface, while the other contact is an STM tip.   

For both setups, we initially relax the molecule and the two closest Au layers are using a double-zeta with polarization (dzp) atomic basis set~\cite{gpaw-lcao} using the grid-based projector augmented wave method GPAW~\cite{gpaw}  with the Perdew-Burke-Ernzerhof GGA exchange-correlation functional~\cite{PBE}. In the relaxed configuration, the S-atoms bind to Au at a bridge site slightly shifted toward the hollow site.  Following the standard DFT-Landauer approach as described in  Ref.~\onlinecite{strange08}, we calculate the zero-bias transmission function, $T(E)$. The low-bias conductance can finally be obtained from the Landauer formula, $G=(2e^2/h)\,T(E_F)$, where $E_F$ is the Fermi energy.

Figure \ref{thio_trans} (c) shows the electronic transmission function for the AQ (dashed black) and HQ (solid red) molecular bridges show in panel (a).  There is a strikingly large difference of six orders of magnitude in the transmission around the Fermi energy. While the HQ has a relatively large transmission of around $0.01$, the AQ displays a clear transmission anti-resonance with a minimum transmission of $10^{-8}$. 

Figure \ref{thio_trans} (d) shows the transmission functions of the HQ and AQ in the STM setup (Figure \ref{thio_trans} (b)). The transmission values are generally lower by a factor 100 compared to the contact in Figure \ref{thio_trans} (a). This is due to the weak tunneling contact between the molecule and the tip. We again observe a clear transmission dip for the AQ but not for the HQ. 
Both the transmission peaks and dip are shifted toward lower energies in the STM setup, and the antiresonance is no longer right at the Fermi level. We attribute this to the different coupling to the Au contact which cause different shifts of the molecular levels. It might still be possible that even the transmission antiresonance away from the Fermi level can be observed experimentally by tuning the molecular levels with the electrochemical gate voltage.
The qualitative agreement between the two different contact situations show that the interference feature in the AQ is an intrinsic feature of the molecule, which is only quantitatively affected by the coupling to the metal. The large on/off ratio should therefore be a robust feature of the AQ-HQ switch. In a recent STM conductance measurement on quinone-oligo(phenylene vinylene) (Q-OPV) molecules in an electrochemically controlled environment~\cite{TsoiNano2008}, it was found that the reduced HQ-OPV molecule has a 40 times larger conductance than the Q-OPV. This results is in qualitative agreement with our finding that HQ has a larger conductance than AQ. 

It is well known that DFT in general does not provide an accurate description of molecular energy gaps and level alignment at metallic surfaces\cite{Toher2005,juanma09}. In order to verify that the interference effect is not sensitive to changes in the molecular level positions we have changed ``by hand'' the molecular orbitals using a scissors operator technique similar to that described in Ref. \cite{Mowbray2008}. Shifting the energies of the occupied states down by 1 eV and the energies of the unoccpied states up by 1 eV slightly shifts the transmission minimum of the AQ, but does not change the overall picture. The large on/off ratio should thus be a rather robust feature of the AQ-HQ switch. 

In the remainder of the paper we analyze the AQ-HQ switch in terms of two simplifying models that offers complementary insights to the underlying physics of the interference effects.

\section{H\"uckel $\pi$-model}
\begin{figure}[htb!]
\begin{center}	
\includegraphics[width=\columnwidth,angle=0]{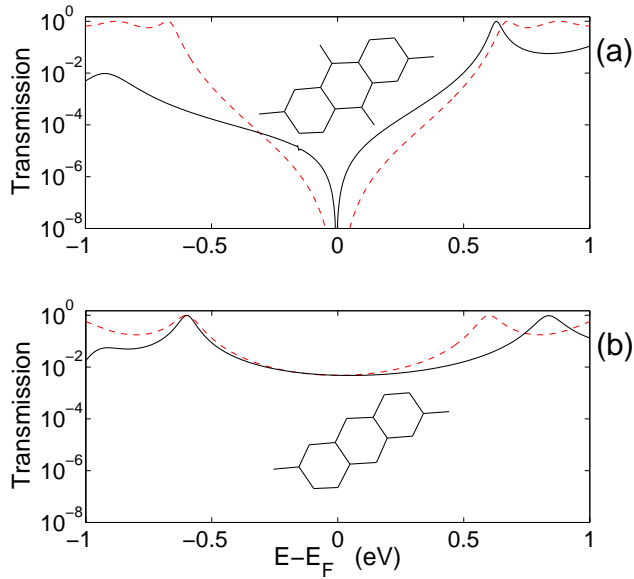}
\end{center}	
\caption{Transmission of the AQ (a) and HQ (b) molecules calculated with a H\"uckel $\pi$-model (dashed red) and compared with the DFT-GF results (solid black). The insets show the structure used in the H\"uckel model}
\label{pi-model}
\end{figure}

In order to gain insight in the physics underlying the interference effects, we describe the AQ/HQ molecule with a H\"uckel $\pi$-electron model, in which there is one $p_z$-orbital at every carbon atom. We assume only nearest neighbour interaction with a hopping parameter, $t=-3\,$eV, and constant on-site energy $\varepsilon_0=0\,$eV.  In the oxidized AQ state, we further assume that the oxygens contribute with one $p_z$ orbital each with the same on-site and hopping parameters as the C atoms. In the reduced HQ state this extra electron will be passivated by the hydrogens, and does not contribute to the $\pi$ system. 
The schematics of the central parts of the AQ and HQ molecules in the $\pi$-model are shown as insets in Fig. \ref{pi-model} (a) and (b), respectively. The transmission in the $\pi$-model is calculated using a wide band limit approximation for the electrode self-energies, which adds an imaginary part, $i\gamma$, to the on-site energy of the C-atoms connected to the sulfurs. The sulfur atoms themself are not included in the model. 

Figure \ref{pi-model} shows the full DFT transmission (solid black) together with the $\pi$-model transmission (dashed red) for the AQ (a) and HQ (b) molecules. Both the transmission anti-resonance in the AQ and the relatively large transmission for the HQ found in the DFT-GF calculations are well reproduced by the $\pi$-model. The good agreement between the DFT-GF and the $\pi$-model results is not obvious as it might be expected that both the on-site energy of and the hopping to the oxygens are quite different from the values for pure carbon. However, it is encouraging that the simple $\pi$-model reproduce the DFT results so well. The almost exact coexistence of the transmission minimum at the Fermi energy, might not be a general result, since the exact position of the DFT transmission minimum is sensitive to the bonding geometries, as illustrated by the differences in transmission between the two setups in Fig. \ref{thio_trans}. 

The results of the $\pi$-model shows that switching from HQ to AQ can be thought of as effectively adding an orbital to the $\pi$-system. An electron in AQ travelling from left to right can either do it directly or via the extra side orbital, with the two pathways intefering destructively at the Fermi energy. In the HQ, on the other hand, there is only the direct path and no interference effects should be expected.

\section{Localized molecular orbitals}
\begin{figure}[htb!]
\begin{center}	
\includegraphics[width=\columnwidth,angle=0]{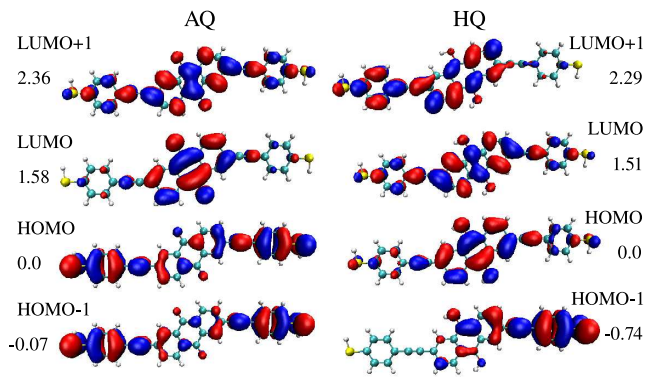}
 \end{center}
\caption{Molecular eigenstates of the isolated AQ and HQ, respectively. Note that the HOMO and LUMO for the HQ have similar shapes as the LUMO and LUMO+1 for the AQ. The effect of the two extra H atoms in the HQ is thus merrily to occupy one extra orbital without significantly changing the orbital shapes. The energies of the eigenstates relative to the HOMO levels are given in units of eV.}
\label{free_MO}
\end{figure}

While the H\"uckel model succesfully explains the interference effects in terms of an extra $\pi$-orbital in the oxidized AQ state, it is desirable to understand the interference effects in terms of the frontier orbitals. To this end, we develop in this section a simple and general method for analyzing quantum interference in molecular junctions using the AQ/HQ switch in Figure \ref{thio_trans} (a) as a specific example. In the physically relevant energy range around the Fermi level, the transmission through a molecular junction is governed by the frontier molecular orbitals (MOs) and their overlap with the electrode states.  Often a qualitatively correct description can be obtained by considering only the single molecular level closest to Fermi energy, i.e. either the highest occupied molecular orbital (HOMO) or the lowest unoccupied molecular orbital (LUMO). Assuming a constant density of states (DOS) in the electrodes (the so-called wide band limit), the transmission function of a single-level model becomes a Lorentzian centered at the energy of the MO and with a width, $\Gamma$ determined by the electrode DOS and the spatial overlap of the MO with the states in the electrode. The frontier MOs of the free (i.e. not coupled to gold) AQ and HQ are shown in Figure \ref{free_MO}, together with the level energies.

\begin{figure*}[htb!]
\begin{center}		
\includegraphics[width=\textwidth,angle=0]{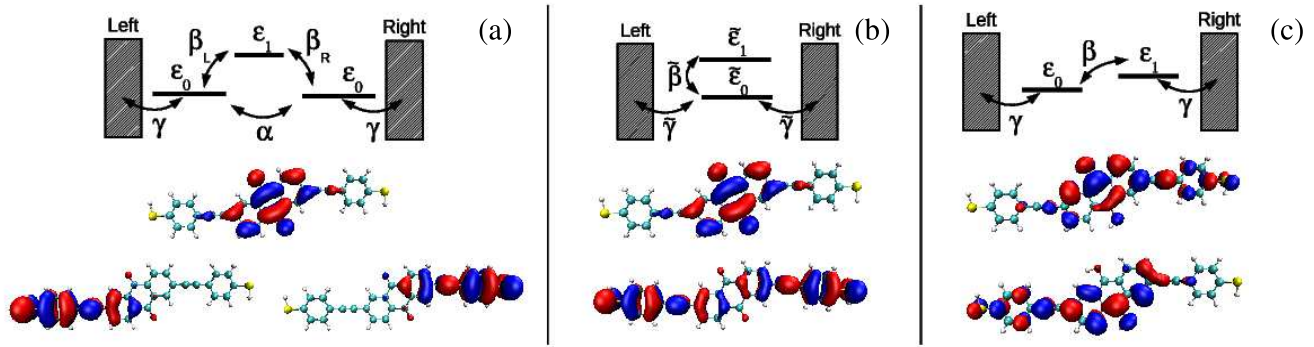}
\end{center}
\caption{Localized molecular orbitals (LMOs) and the corresponding TB models. The LMOs have been constructed from (a) the HOMO-1,HOMO, and LUMO of AQ (b) the HOMO and LUMO of AQ, and (c) the HOMO and LUMO of HQ. TB models (a) and (b) offers the opportunity for destructive interference and transmission anti-resonances due the multiple paths through the molecule.}
\label{thio_wannier}
\end{figure*}

In Figure \ref{thio_trans} (c) the transmission function of HQ can approximately be described by a sum of two Lorentzians centered at the HOMO and LUMO energies, respectively. In contrast this is not the case for AQ. The pronounced difference in the transmission properties of AQ and HQ is difficult to detect from the shape and energies of the frontier MOs (see Figure \ref{free_MO}). The difference, however, becomes clear if we transform the frontier MOs, $\psi_m$, into localized molecular orbitals (LMOs), $\phi_n$:
\begin{equation}
 \phi_n = \sum_m U_{mn}\psi_m
\end{equation}
The expansion coefficients are determined following the standard procedure for constructing maximally localized Wannier functions (for solids) or maximally localized molecular orbitals (for molecules)~\cite{vanderbilt_WF,thygesen_WF}. We thus seek to minimize the spread of the LMOs measured by the sum of their second moments:
\begin{equation}
 S=\sum_n\left(\langle \phi_n|r^2|\phi_n\rangle - \langle \phi_n|\mathbf{r}|\phi_n\rangle^2\right). \label{spread}
\end{equation}
 
The mapping from the molecular eigenstates ($\psi_m)$ to the LMOs ($\phi_n$) described by the unitary matrix, $\Ub$, is uniquely determined by the requirement that the resulting LMOs should be maximally localized as defined by Eq. \eqref{spread}. Usually, one maps the entire set of occupied MOs into an equivalent set of LMOs. In the present case, however, we are interested in describing the physics in the vicinity of the Fermi level and thus we transform only the MOs closest to $E_F$. Using the LMOs as basis states we can construct an effective TB-like Hamiltonian for the molecule.  As an example Figure \ref{thio_wannier}(a) shows the LMOs obtained from the HOMO-1, HOMO, and LUMO of the AQ. The corresponding TB Hamiltonian is obtained as 
\begin{eqnarray}  
\mathbf{H}&=&\Ub^T\,{\rm diag}(\varepsilon_{\hbox{\rm \tiny  H-1}},\varepsilon_{\hbox{\rm \tiny  H}},\varepsilon_{\hbox{\rm \tiny  L}})\,\Ub 
=\begin{pmatrix}
    \varepsilon_0 & \beta_L & \alpha \\
    \beta_L & \varepsilon_1 &  \beta_R \\
    \alpha & \beta_R & \varepsilon_0 \\
   \end{pmatrix} \label{3siteH},
\end{eqnarray}
where $\varepsilon_{\text{H}-1},\varepsilon_{\text{H}},\varepsilon_{\text{L}}$ denote the molecular energy levels. In the considered case the TB Hamiltonian has certain symmetries which derives from symmetries of the LMOs: The two bottom LMOs of Figure \ref{thio_wannier}(a) both have on-site energy $\varepsilon_0=0.07\,$eV (relative to the AQ HOMO level) but are localized on the left and right parts of the molecule, respectively. The topmost LMO has on-site energy $\varepsilon_1=1.56\,$eV and is localized at the center of the molecule with a vanishing weight at the terminating S atoms. Assuming a wide band approximation, the lead self-energies only adds an imaginary part to  the left and right on-site energies, $\varepsilon_0\rightarrow \varepsilon_0+i\gamma$. This express the fact that the molecule is only coupled to the leads via the two LMOs located at the ends of the molecule.  The topology of the effective three-site TB model is sketched in the upper panel of Figure \ref{thio_wannier} (a). The hopping parameters are $\alpha=0.07\,$eV and $\beta_L=\beta_R=0.25\,$eV. It is intuitively clear that the two pathways through the molecule, i.e. hopping from the left to the right LMO either directly or via the central LMO, can lead to interference effects in the transmission.

The LMOs and corresponding TB model obtained through the procedure
described above depends on the initial set of MOs used for the
mapping. In Figure \ref{thio_wannier}(b) we show the LMOs and the
corresponding TB topology obtained when the mapping is performed from
the HOMO and LUMO of the AQ. The obtained TB parameters are $\tilde{\varepsilon}_0=0.15\,$eV, $\tilde{\varepsilon}_1=1.56\,$eV, $\tilde{\beta}=0.35\,$eV. As for the three-state case one can
anticipate interference effects due to transmission though two
different pathways -- either direct hopping or hopping via the LMO
localized at the center.

\begin{figure}[htb!]
\begin{center}
\includegraphics[width=\columnwidth,angle=0]{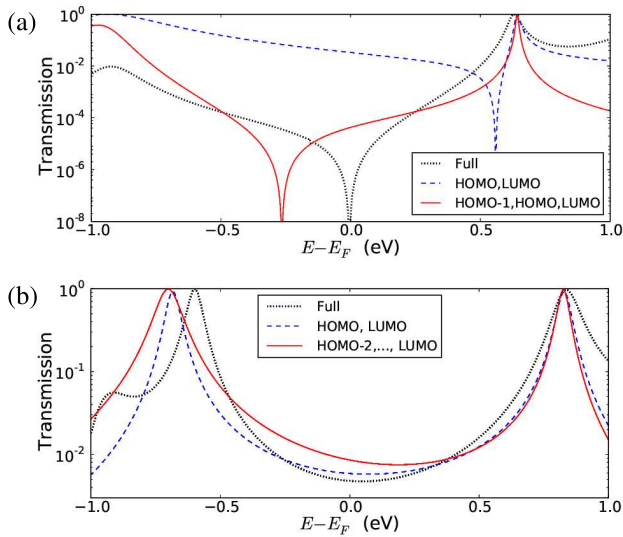}
\end{center}
\caption{(a) Transmission function of the AQ coupled to gold electrodes calculated using the DFT-GF method (dotted black) and for the two-site (dashed blue) and three-site (solid red) TB models in Figure \ref{thio_wannier} (b) and (c). (b): DFT-GF  transmission (dotted black) for the HQ together with TB transmission  for the two-site model (dashed blue) in Figure \ref{thio_wannier} (c) and for a four   site-model including two extra occupied MOs (solid red).} 
\label{trans_wannier}
\end{figure}

Figure \ref{trans_wannier} (a) shows the transmission of the AQ (on a
log scale) calculated using the full DFT-GF setup (dotted black),
which is the same as shown in Figure \ref{thio_trans} (c). Also shown are
the transmissions obtained from the three- and two-site TB models in
Figure \ref{thio_wannier} (a)-(b). As we do not have any reference to
the electrode Fermi energy in the TB models, we have aligned the LUMO
transmission peak (at $E=0.6\,$eV) of the TB transmissions to the
DFT-GF result. Evidently, the three-site model (solid red) qualitatively
reproduces the transmission anti-resonance and the overall shape of
the transmission function is better reproduced by the three-site model
than the two-site model (dashed blue), which in general yields a too large
transmission and the transmission anti-resonance is too high in
energy. It can readily be found that the two-site model has a
transmission zero at energy $E=\tilde{\varepsilon}_1$, whereas the
three-site model has the anti-resonance at
$E=\varepsilon_1-\beta_L\beta_R/\alpha$. Note that in both models the
anti-resonance energy is independent of the on-site energies,
$\varepsilon_0$ and $\tilde{\varepsilon}_0$, and of the LMOs coupling to
the electrodes. It is not surprising that the three-site model gives a better description of
the full calculation as the energies and wavefunctions of the HOMO and HOMO-1 are very similar, see  Figure \ref{free_MO}.
We note that the quantitative differences between the full and
model calculations are due to (i) neglect of all but a few MOs in the
construction of the LMOs, and (ii) the wide band approximation.

We now turn to the HQ molecule. Figure \ref{thio_wannier}(c) shows the LMOs constructed from the HOMO and LUMO orbitals, and the corresponding TB model. The on-site energies are $\varepsilon_0=0.55\,$eV, $\varepsilon_1=0.97\,$eV (relative to the HQ HOMO level), and $\beta=-0.73\,$eV. The topology of the HQ two level model is qualitatively different from the AQ topologies in that there is only
one possible pathway through the molecule and thus interference effects are not anticipated. Figure \ref{trans_wannier}(b) compares the
transmission through the HQ obtained from the full calculation (dotted black), the two-site TB model in Fig. \ref{thio_wannier}(c) (dashed blue), and a four-site model including also the HOMO-2 and HOMO-1 (solid red). Both TB models reproduce the full result, and neither of them shows interference around the Fermi level. The four-site model gives an anti-resonance at energies below the HOMO level which correspond to the anti-resonance in the AQ.

\subsection{Application of LMOs to a different molecule}

\begin{figure}[htb!]
\begin{center}
 \includegraphics[width=0.9\columnwidth,angle=0]{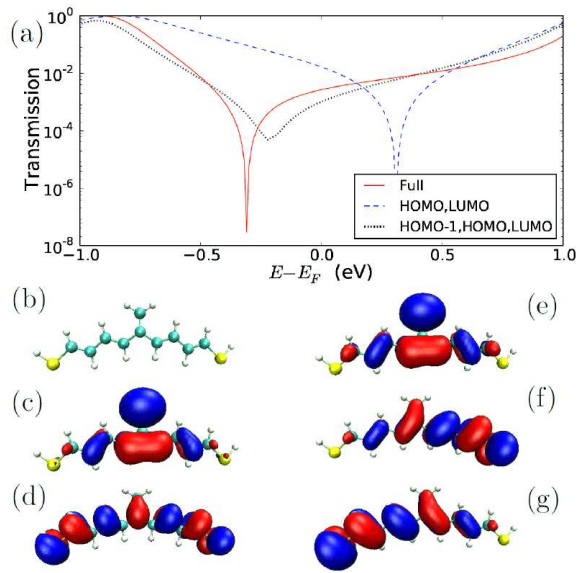}
\end{center}
\caption{(a): Transmission through the molecule shown in panel (b) calculated with the DFT-GF method with the molecule coupled to gold electrodes (solid red) and within the wideband approximation using the the two- (dashed blue) and three-site (dotted black) models in Figure \ref{thio_wannier} (b) - (c). The corresponding LMOs are shown in panel (c)-(d) and (e)-(g) for the two- and three site models, respectively. }
\label{trans_solomon}
\end{figure}

To further validate the simple model calculations using LMOs of the
isolated molecules, we consider the cross-conjugated molecule shown in
Figure \ref{trans_solomon} (b). The transmission properties of similar
molecules have previously been analyzed~\cite{Solomon2008} and clear
transmission anti-resonances were observed, similar to the AQ
transmission shown above. In Figure \ref{trans_solomon} (a) we compare
the transmission through the cross-conjugated molecule attached to Au(111)
electrodes as above and calculated with DFT-GF (dotted black), and
with two TB models. Our calculated transmission agrees with the
results in Ref.~\cite{Solomon2008}. The three-site model (solid red)
gives again a qualitatively correct description of the transmission
anti-resonance. The corresponding LMOs, constructed from the
HOMO-1,HOMO, and LUMO, are shown in Figure \ref{trans_solomon} (e)-(g). 
The two-site orbitals are shown in Figure \ref{trans_solomon} (c)-(d). As for the AQ, the two-site model yields a transmission anti-resonance higher in energy than both the three-site model and the full calculation.  The good agreement between the three-site TB model and the full DFT-GF calculation further illustrates that
the simple minimal TB model provides a transparent and qualitative
correct description of the full transport problem.  The molecular level energies for the free cross-conjugated molecules are $\varepsilon_{\mbox{\tiny{HOMO-1}}}=-0.4\,$eV, $\varepsilon_{\mbox{\tiny{HOMO}}}=0.0\,$eV, $\varepsilon_{\mbox{\tiny{LUMO}}}=2.0\,$eV, relative to the HOMO level. Contrary to the AQ, the HOMO-1 and HOMO levels are clearly separated in energy, and the better results obtained with the three-site model including the HOMO-1 orbital indicates that quantum interference effects in molecules more generally involves the HOMO-1, HOMO, and LUMO orbitals and not only the HOMO and LUMO, as might be anticipated.

\section{Summary}
In summary, we have performed \textit{ab initio} transport calculations for a recently proposed anthraquinone based electrochemical molecular switch~\cite{DijkOrgLett2006}. Due to destructive quantum interference effects the conductance in the reduced ('off') state is dramatically suppressed by several orders of magnitude as compared to the oxidized ('on') state. The desctructive interference were explained within a simple H\"uckel $\pi$-model by the addition of two extra $p_z$ orbitals on the oxygen atoms in the oxidized anthraquinone state. The interference effects were further rationalized by mapping the full transport problem to an effective minimal TB model obtained via a transformation of the frontier molecular orbitals into localized molecular orbitals. The approach can be used to categorize any given molecular junction according to the topology of the effective TB model and offers a transparent and fast method for computational screening for molecules exhibiting interference features.

\textbf{Acknowledgement} We are grateful to Sense Jan van der Molen
for inspiring discussions. The authors acknowledge support from FTP grant nr. 274-08-0408
and The Danish Center for Scientific Computing. The center
for Atomic-scale Materials Design (CAMD) is supported by the Lundbeck
Foundation. The Center for Individual
Nanoparticle Functionality (CINF) is sponsored by the
Danish National Research Foundation.


\end{document}